\documentstyle[psfig,preprint,aps,prl]{revtex}
\begin{document}
\firstfigfalse
\title{Notes on the static dielectric response function in the density functional
theory}
\author{I.I. Mazin and R.E. Cohen}
\address{Carnegie Institution of Washington and Center for High Pressure Research,\\
5251 Broad Branch Rd., Washington, DC 20015}
\maketitle

\begin{abstract}
We discuss several aspects of the dielectric response theory application to
the density functional theory. This field has been an unceasing source of 
confusion during several decades. The most frequent reasons for this 
confusion are (a) uncritical tranfer of the results, especially regarding
so-called local field corrections, obtained in many-body perturbation 
theory onto density functional theory, and (b) mixing up the statements true
for the exact density functional theory with those applicable to the 
local density approximation only. In these notes we try to draw an 
appropriate lines between those theories. We also discuss a newly introduced
(X. Gonze, Ph. Ghosez, and R.W. Godby, Phys. Rev. Lett., {\bf 74}%
, 4035, 1995) "polarization+density functional" and show that within a
given (e.g., local density) approximation to the exchange-correlation energy
the Gonze et al approach is exactly equivalent to the conventional one.
\end{abstract}

Even now, 30 years after the discovery of the density functional theory (DFT)%
\cite{DFT} there is still considerable confusion about applicability of DFT
to static dielectric response\cite{1}. On the one hand, there is seemingly
no room for questioning the validity of the DFT in this respect: The
Hohenberg-Kohn theorem, which is the basis of DFT, is a mathematical
theorem, rigorously proven, that states that the total energy of the ground
state of a many-electron system is a unique functional of its electronic
density, and the density-density response function, defined as 
\[
\chi ({\bf r,r}^{\prime })=\delta ^2E/\delta \rho ({\bf r)}\delta \rho ({\bf %
r}^{\prime }),
\]
is also a unique functional of the electronic density. Surprisingly, this
simple theorem is not readily digested by everybody. The reasons for
confusion are several: First, one commonly confuses the exact DFT with its
local, approximate version (LDA). As we shall discuss below, mixing up these
two notions is much more dangerous when dealing with dielectric response than
with the total energy itself. Second, to have this theorem satisfied, one
has to take into account properly exchange-correlation corrections to the
dielectric susceptibility --- and be aware that those corrections are
functionally different in DFT, and, say, in the many-body perturbation
theory or in the Fermi-liquid theory (needless to say that all observable
quantity are the same in any of these theories). Third, the 
notorious failure of
LDA-DFT to produce the
correct band gap (well understood by now, thanks to
seminal works of Levy and Perdew \cite{LP} and Sham and Schl\"{u}ter\cite{SS}%
), combined with the conventional wisdom that the dielectric gap determines
the response, forces ingenuous researchers to question the formal
applicability of the DFT to response functions calculations. Finally, in the
last decades accurate LDA-DFT calculations of the dielectric constant of
semiconductors have been performed\cite{calc}, which were usually in error
by 10-15\%. Incredible success of the LDA calculations of structural
properties (a few per cent) made people imply indirectly that similar
accuracy is attainable in other ground-state properties calculations. When
it turned out to be not true, a suspicion rose that something may be
principally wrong with the approach (in reality, of course, 10-15\% accuracy
is excellent for such a simple approximation as LDA). In particular, the
Hohenberg-Kohn theorem itself has been questioned\cite{1}.
 The goal of the current
notes is not to produce a new physical result but to help, especially
newcomers in the field, to avoid confusion and misleading of incorrect
claims spread around in the literature.

One of the source of confusion is the fact that exchange-correlation local
field (XCLF) corrections are formally different in all three theoretical
approaches to the dielectric response of many-electron systems. The most
traditional one is the many-body perturbation theory, some times referred to
as Green function theory.
In this theory one-electron excitations are poles of the Green function,
$G^{-1}({\bf r,r}^{\prime },\omega )$
determined by the Dyson equation\cite{AGD}, 
\[
G^{-1}({\bf r,r}^{\prime },\omega )=G_0^{-1}({\bf r,r}^{\prime },\omega
)-\Sigma ({\bf r,r}^{\prime },\omega )
\]
where $\Sigma $ is the self-energy operator, and
$G^{-1}({\bf r,r}^{\prime },\omega )$ is the Green function of free
electrons. An alternative form for the Dyson
equation is 
\begin{equation}
H_0({\bf r})\psi _n({\bf r},E_n)+\int \Sigma ({\bf r,r}^{\prime },E_n)\psi
_n({\bf r}^{\prime },E_n)=E_n\psi _n({\bf r},E_n),  \label{GF}
\end{equation}
where $H_0$ is the one-electron Hamiltonian, $\psi _n({\bf r},E_n)$ is the
Green function amplitude and $E_n$ is the corresponding pole. $E_n
$ are in general complex. In the random phase approximation the dielectric
susceptibility of the system can be written in terms of the bare
polarization operator as 
\[
\chi _{RPA}^{GF}(1,2)=[1-V_C(1-3)\Pi _0^{GF}(3,4)]^{-1}\Pi _0^{GF}(4,2),
\]
where we introduced superscript $GF$ for the Green functions theory, and
integration over the space coordinates is implicitly implied. Here
1, 2, etc are short for ${\bf r_1}$, ${\bf r_2}$, etc. $\Pi_0$ is
the bare polarization operator. Symbolically
one can write this equation  as 
\begin{equation}
\chi _{RPA}^{GF}=\frac{\Pi _0^{GF}}{1-V_C\Pi _0^{GF}},  \label{GF-RPA}
\end{equation}
and we shall use such notation throughout the paper, when unambiguous. Going
beyond RPA demands replacing $\Pi _0^{GF}$ by renormalized (full
irreducible) polarization operator $\Pi ^{GF}.$ The XCLF, as
introduced by Hubbard in 1958 \cite{Hubbard}, assumes that $\Pi ^{GF}$ can
be approximately written as 
\begin{equation}
\Pi ^{GF}\approx \Pi _0^{GF}/(1-I\Pi _0^{GF}),  \label{GF-Pi}
\end{equation}
where exchange-correlation
 interaction $I({\bf r,r}^{\prime })$ is negative (as well as $\Pi
_0^{GF})$, and thus the total susceptibility is enhanced compared to RPA: 
\begin{equation}
\chi \approx \frac{\Pi _0^{GF}}{1-(V_C+I)\Pi _0^{GF}}=\frac{\chi _{RPA}^{GF}%
}{1-I\chi _{RPA}^{GF}}.  \label{GF-Chi}
\end{equation}

An important fact is that the difference between $\Pi ^{GF}$ and $\Pi _0^{GF}
$ comes first of all from exchange processes, namely 
\begin{figure}[tbp]
\centerline{\psfig{file=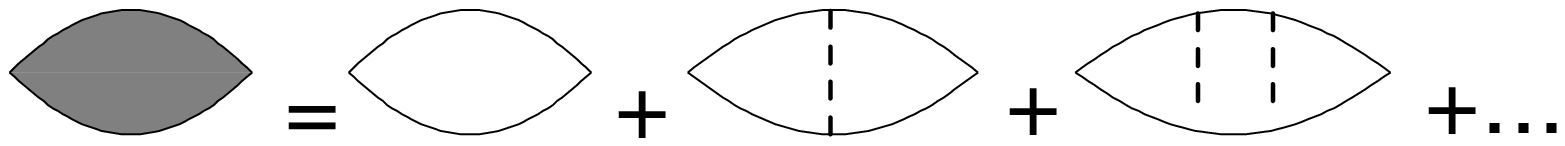,width=0.95\linewidth}}
\end{figure}

In semiconductors, the exchange interaction remains long-ranged. In other
words, when this series is approximated by Eq. \ref{GF-Pi}, 
\begin{figure}[tbp]
\centerline{\psfig{file=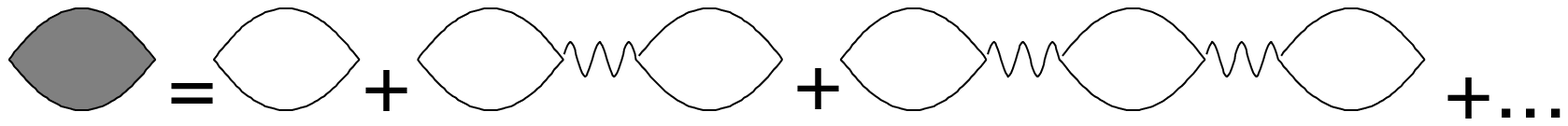,width=0.95\linewidth}}
\end{figure}
the corresponding effective interaction $I$ (wavy line)
diverges in reciprocal space as $%
1/q^2$. As we shall see below, it is not the case in LDA.

Formula (\ref{GF-Chi}) looks very similar to the expression for the
susceptibility in the Fermi-liquid theory 
\begin{equation}
\chi =\frac{\chi _{RPA}^{FL}}{1-f\chi _{RPA}^{FL}}.  \label{FL-Chi}
\end{equation}
However, while the left-hand sides of Eqs. \ref{GF-Chi} and \ref{FL-Chi} are
the same, $\chi _{RPA}^{FL}$ differs from $\chi _{RPA}^{GF},$ as it usually
defined in the many-body theory, in the sense that it is calculated with
exact one-electron excitation spectrum, given by the poles of the full Green
function (\ref{GF}), while $\chi _{RPA}^{GF}$ assumes bare Green functions.
For instance, in the theory of the homogeneous electron gas $\chi _{RPA}^{GF}$
is just the Lindhardt susceptibility. However, in practical calculations for
semiconductors, for instance in the GW approximation, it is common to call
``RPA'' susceptibility the quantity calculated with fully renormalized
one-electron spectrum, that is, $\chi _{RPA}^{FL}$. Correspondingly, effective
interaction $f({\bf r,r}^{\prime }),$ the effective Landau interaction, is
the analog of Coulomb + Hubbard interaction in the Hubbard approximation.

The most common way to calculate the dielectric function is related to the
density functional theory. To refresh readers' memory, let us remind the
basic equation of this theory: 
\[
H_{DFT}({\bf r})\varphi _n({\bf r})\equiv H_0({\bf r})\varphi _n({\bf r}%
)+V_{KS}\varphi _n({\bf r}^{\prime })=\varepsilon _n\varphi _n({\bf r}),
\]
where  the wave functions $\varphi $ and the
(real) eigenvalues $\varepsilon $ are for a fictitious system of
non-interacting fermions with  the same density as
the electronic system in question. The above equation is a tool to find this
spectrum, and the Kohn-Sham potential $V_{KS}$ is defined as 
\[
V_{KS}({\bf r})=\frac{\delta E_{int}[\rho ({\bf r})]}{\delta \rho ({\bf r})}.
\]
$E_{int}$ is a unique functional of the total density, and the total energy
of the electronic system is, by definition, $\sum_{\varepsilon <\mu }\langle
\varphi _i|-\frac{{\bf \nabla }^2}{2m}|\varphi _i\rangle +E_{int}[\rho ({\bf %
r})]$, where $\mu $ is the chemical potential. Since the dielectric
susceptibility can be defined entirely in terms of the total energy and
total density, 
\begin{equation}
\chi ({\bf r,r}^{\prime })=-\frac{\delta E_{tot}[\rho ]}{\delta \rho ({\bf r}%
)\delta \rho ({\bf r}^{\prime })},  \label{Echi}
\end{equation}
it is also a unique density functional. Unlike the Green functions theory,
DFT allows a formally exact expression for $\chi $, which follows directly
from Eq. \ref{Echi} and looks similar to the {\it approximate} Eq.\ref
{GF-Chi} in the Green functions theory: 
\begin{equation}
\chi =\frac{\chi _{RPA}^{DFT}}{1-I_{xc}\chi _{RPA}^{DFT}},  \label{DFT-Chi}
\end{equation}
where now by definition 
\[
I_{xc}({\bf r,r}^{\prime })=\frac{\delta ^2E_{xc}[\rho ]}{\delta \rho ({\bf r%
})\delta \rho ({\bf r}^{\prime })}=\frac{\delta V_{xc}[\rho ]({\bf r)}}{%
\delta \rho ({\bf r}^{\prime })}.
\]

There is principal difference between $I_{xc}$ behavior in LDA and exact
DFT: In the former, $I_{xc}({\bf r,r}^{\prime })\propto \delta ({\bf r-r}%
^{\prime })$, while in the latter $I_{xc}({\bf r,r}^{\prime })$ may be of
arbitrary long range. A good example (which was communicated to the authors
by O. Gunnarsson) is exact DFT for an insulator: adding one electron to
arbitrary large insulator induce uniform shift of the exchange-correlation
potential for the whole system (so-called ``density-derivative
discontinuity'', Refs.\cite{LP,SS}), thus making $I_{xc}({\bf r,r}^{\prime
}) $ of infinite range. Since the density-derivative discontinuity is always 
non-zero (because of the exchange-correlation part of kinetic energy, see
Ref. \cite{LP}), that means that in the exact DFT $I_{xc}({\bf q})$ diverges
at $q\rightarrow 0$ as $\delta (q)$. Nevertheless, the magnitude of the
divergency may be arbitrary small. In fact, 
it appears that the DFT expression (%
\ref{DFT-Chi}), if the spacial dependence is properly dealt with, yields
good results (with 10-15\% accuracy) even in LDA \cite{calc}. Again, it is
instructive to compare the longe-distance behavior of $I_{xc}$ in GF theory,
in LDA and in the exact DFT:  
in GF theory $I_{xc}$ diverges, in insulators, 
at $q\rightarrow 0$ as $1/q^2$; this statement is not necesserily true
in exact DFT, where the most divergent term is just a $\delta$-function
and nothing can be said rigorously about the next terms. Finally, in LDA 
$I_{xc}$ remains constant at $q\rightarrow 0$\footnote{One can hear
occasionally statements that $I_{xc}^{DFT}$ should diverge as $1/q^2$ at
at small $q$'s; we are not aware of any proof of this statement which would
not indirectly use an unproven parallel with $I_{xc}^{GF}$.}

The spatial dependence is rather important. Proper treatment of the spacial
dependence includes Umklapp processes, 
\begin{eqnarray}
1/\epsilon (q) &=&[\epsilon _{{\bf q+G,q+G}^{\prime }}]_{00}^{-1}\,
\label{LF} \\
\epsilon _{{\bf q+G,q+G}^{\prime }}^{-1} &=&\delta _{{\bf GG}^{\prime }}+%
{\rm V}_C\cdot \Pi _0^{DFT}\cdot \left[ \delta _{{\bf GG}^{\prime }}-\left( 
{\rm V}_C{\rm +I}_{xc}\right) \Pi \right] ^{-1}  \nonumber
\end{eqnarray}
where ${\rm V}_C({\bf q+G,q+G}^{\prime })=4\pi e^2\delta _{{\bf GG}^{\prime
}}/|{\bf q+G|}^2,$ and ${\rm I}_{xc}$ is the Fourier transform of $I_{xc}(%
{\bf r,r}^{\prime }),$ and the tensor dot-products are taken in the
right-hand side. All terms here which originate from $G,G^{\prime }\neq 0$
are called ``local field corrections''. An elegant approach which is
mathematically equivalent with Eq.\ref{LF}, and is often called ``the
Sternheimer equation'', avoids using polarization operators explicitly, but
instead deals directly with the change of the density (see, e.g. Ref. \cite
{Gianozzi}, Eqs. 12-16: 
\begin{eqnarray}
\delta \rho ({\bf q+G)} &=&4\sum_{{\bf k},n\in occ}\langle \psi _{n{\bf k}%
}|e^{-i({\bf q+G)r}}P_c|\delta \psi _{n{\bf k+,q}}\rangle   \label{Stern} \\
(\varepsilon _{n{\bf k}}-H_{DFT} &&)|\delta \psi _{n{\bf k+q}}\rangle
=P_c\delta H_{DFT}|\psi _{n{\bf k}}\rangle ,  \nonumber
\end{eqnarray}
where $P_c$ is the projector operator on the conduction (unoccupied) bands.
These equations are formally equivalent to Eqs.\ref{LF} and thus
are exact  within the DFT. Furthermore, if one uses the same (approximate)
exchange-correlation energy functional, for example LDA, both sets of
 equations (\ref{LF}) and (\ref{Stern}) should yield the same result.

Much of confusion was raised by the fact that the Hohenberg-Kohn theorem
applies to the {\it total} electron density. That is, upon applying a long wave ($%
q\rightarrow 0)$ perturbation, the total energy depends not only on the {\it %
periodic} part of the density change, $\sum_{{\bf G}\neq 0}\delta \rho ({\bf %
q+G),}$ but also on the long wave part, $\delta \rho ({\bf q+0).}$ Despite
this fact, neither Eqs. \ref{LF} nor Eqs. \ref{Stern} demand supercell
calculations with periodicity $1/q$. The reason is that the original,
unperturbed system, which is periodic, bears all information (in linear
regime) about the properties of the system, perturbed by any external field
(as long as it weak enough), including long-range perturbation. It was
claimed recently by Gonze et al \cite{1} that Eqs. \ref{Stern} are
incomplete {\it because the do not treat the long-wave part of the density
properly.} This is of course a fallacy. Interestingly, trying to overcome
non-existent incorrectness of Eqs. \ref{Stern}, Gonze et al derived another
equation, but failed to realized that it was mathematically equivalent to
Eqs. \ref{Stern}. This is worth elaborating.

Gonze et al suggested in Ref.\cite{1} to use a functional of two functions:
the periodic electron density, $\rho _{per}=\sum_{{\bf G}\neq 0}\rho ({\bf %
q+G)}e^{i({\bf q+G)r}}$, and macroscopic polarization ${\bf P(q).}$ The
latter is uniquely related to the macroscopic density perturbation, $\rho
_{mac}=\rho ({\bf q)}e^{i{\bf qr}}\approx i({\bf qr)}\rho ({\bf q)}$ for $%
q\rightarrow 0,$ so ${\bf P}$ and $\rho _p$ together uniquely define the total
density and therefore the total energy. Obviously, one has full freedom to
chose with which variables to work, periodic density plus macroscopic 
density, or periodic density plus macroscopic polarization.
Having chosen the second scheme, Gonze et al
derived equations where instead of the change of the total exchange correlation
potential in $\delta H_{DFT}$ they used only periodic part of this
potential. Their final expression corresponded to substitution of $P_c\delta
V_{xc}^{tot}|\psi _{n{\bf k}}\rangle $ in Eqs. \ref{Stern} by 
\begin{equation}
P_c\delta V_{xc}^{per}|\psi _{n{\bf k}}\rangle +[\frac{\delta ^2E_{xc}}{%
\delta {\bf P}^2}\delta {\bf P+}\sum_{{\bf G}\neq 0}\frac{\delta ^2E_{xc}}{%
\delta {\bf P}\delta \rho _{{\bf G}}({\bf r)}}\delta \rho _{{\bf G}}({\bf %
r)]}P_c{\bf r}|\psi _{n{\bf k}}\rangle .  \label{GGG}
\end{equation}
In fact, Gonze et al claimed that previous LDA calculations of the 
dielectric response in semiconductors were incorrect, because they had
used Eqs.\ref{Stern} instead of Eqs.\ref{GGG}.
One
can easily show, however, that the two sets of equations are formally
equivalent: From the definition of $I_{xc}$ it
follows that ($\delta ^2E_{xc}/\delta {\bf P}^2)\delta {\bf P=q}I_{xc}({\bf %
q,q)}\delta \rho ({\bf q)}$, and [$\delta ^2E_{xc}/\delta {\bf P}\delta \rho
({\bf q+G})]\delta \rho ({\bf q+G}) = {\bf q}I_{xc}({\bf q,q+G)}\delta \rho (%
{\bf q+G)}$. One can now notice that the part of the Eqs. \ref{Stern} which
depends on the macroscopic component of the exchange-correlation potential, $%
\delta V_{xc}^{mac},$ is nothing else but 
\[
P_c\delta V_{xc}^{mac}({\bf r)}|\psi _{n{\bf k}}\rangle =P_c\sum_{{\bf G}%
}e^{i{\bf qr}}I_{xc}({\bf q,q+G)}\delta \rho ({\bf q+G)}|\psi _{n{\bf k}%
}\rangle , 
\]
which is the same as the second term in Eq. \ref{GGG}. Not surprisingly,
polarization+density functional of Gonze et al appears to be equivalent to
the original Hohenberg-Kohn functional. An unfortunate consequence of this
fact is that the hope that density-polarization functional of Gonze et al
can remedy the
above-mention deficiency of LDA, namely the local character of the
interaction $I_{xc}({\bf r,r}^{\prime })$, is futile. New, more advanced 
approximations to the density funcional are needed to improve the results.
It is possible, although not garanteed,  that these approximations will be 
easier to deal with in the  density-polarization formulation than in the
total-density formulation. A promising routes are generalized density
approximation, much advanced lately, and truly non-local functionals like
weighted-density approxiamtion. In this regard, a recent study by  Dal Corso
et al\cite{GGA}, where a sizable improvement over LDA was found for 
the dielectric constant of silicon, provided that calculations are done
at the same (experimental) lattice parameter.

\end{document}